\documentclass[12pt,preprint]{aastex}


\sfcode`A=1000 \sfcode`B=1000 \sfcode`C=1000 \sfcode`D=1000
\sfcode`E=1000 \sfcode`F=1000 \sfcode`G=1000 \sfcode`H=1000
\sfcode`I=1000 \sfcode`J=1000 \sfcode`K=1000 \sfcode`L=1000
\sfcode`M=1000 \sfcode`N=1000 \sfcode`O=1000 \sfcode`P=1000
\sfcode`Q=1000 \sfcode`R=1000 \sfcode`S=1000 \sfcode`T=1000
\sfcode`U=1000 \sfcode`V=1000 \sfcode`W=1000 \sfcode`X=1000
\sfcode`Y=1000 \sfcode`Z=1000

\shorttitle{IRS Spectroscopy of an ILLBG}
\begin{document}
\title{AEGIS: Infrared Spectroscopy of An Infrared Luminous 
Lyman Break Galaxy at $z=3.01$}
\author{J.-S.~Huang,$\!$\altaffilmark{1}
D.~Rigopoulou,$\!$\altaffilmark{2}
C.~Papovich,$\!$\altaffilmark{3,4}
M. L. N. Ashby,$\!$\altaffilmark{1}
S.~P.~Willner,$\!$\altaffilmark{1}
R.~Ivison,$\!$\altaffilmark{5}
E.~S.~Laird,$\!$\altaffilmark{8}
T.~Webb,$\!$\altaffilmark{6}
G.~Wilson,$\!$\altaffilmark{7}
P.~Barmby,$\!$\altaffilmark{1}
S.~Chapman,$\!$\altaffilmark{9}
C.~Conselice,$\!$\altaffilmark{10}
B.~Mcleod,$\!$\altaffilmark{1}
C.~G.~Shu, $\!$\altaffilmark{11,12}
H.~A.~Smith,$\!$\altaffilmark{1}
E.~Le Floc'h,$\!$\altaffilmark{3}
E.~Egami,$\!$\altaffilmark{3}
C.~A.~N.~Willmer,$\!$\altaffilmark{3}
\& G.~G.~Fazio$\!$\altaffilmark{1}
}

\altaffiltext{1}{Harvard-Smithsonian Center for Astrophysics}
\altaffiltext{2}{Department of Astrophysics, Oxford University}
\altaffiltext{3}{Steward Observatory, University of Arizona}
\altaffiltext{4}{Spitzer Fellow}
\altaffiltext{5}{Royal Observatories, Edinburgh}
\altaffiltext{6}{Department of Physics, McGill University}
\altaffiltext{7}{Spitzer Science Center}
\altaffiltext{8}{UCO/Lick Observatory, Uiniversity of California}
\altaffiltext{9}{California Institute of Technology}
\altaffiltext{10}{Department of Astronomy, Nottingham University}
\altaffiltext{11}{Joint Astrophysics Center, Shanghai Normal University}
\altaffiltext{12}{Shanghai Astronomical Observatory}


\begin{abstract}
  We report the detection of rest--frame 6.2 and 7.7~\micron\ emission
  features arising from Polycyclic Aromatic Hydrocarbons (PAH) in the
  Spitzer/IRS spectrum of an infrared-luminous Lyman break galaxy at
  z=3.01.  This is currently the highest redshift galaxy where these
  PAH emission features have been detected.  The total infrared
  luminosity inferred from  the MIPS 24~\micron\ and radio flux
  density is  2$\times$10$^{13}$ L$_{\odot}$, which qualifies this
  object as a so--called hyperluminous infrared galaxy (HyLIRG).
  However, unlike local HyLIRGs which are generally associated with
  QSO/AGNs and have  weak or absent PAH emission features, this HyLIRG
  has very strong 6.2 and 7.7~\micron\ PAH emission. 
  We argue that intense star
  formation dominates the  infrared emission of this source, although
  we cannot rule out the presence of a deeply obscured AGN.  This LBG
  appears to be a distorted system in the HST ACS F606W and F814W
  images, possibly indicating that a significant merger or interaction
  is driving the large IR luminosity.  
\end{abstract}


\section{Introduction}
  
  Intensive star formation is a very common phenomenon in galaxies at
  high redshifts. Most high redshift galaxies are detected because of
  their high star formation rates, revealed by either the
  rest--frame UV luminosity  in Lyman break galaxies (LBGs, Steidel et
  al. 2003) or at far--infrared wavelengths in, for example,
  submillimeter--selected galaxies (SMGs, Chapman et al. 2003).  It
  was long debated if LBGs and SMGs constituted different coeval
  populations.   Recently \citet{cha05} found that some SMGs have
  similar $u-g$ and $g-R$ colors as LBGs. The Spitzer Space Telescope permits
  exploration of the rest-frame IR properties of LBGs.   For
  example, Reddy et al.\ (2006) studied the 24~\micron\
  properties of $1.5<z<2.5$ LBGs and considered their relation to
  other galaxies at that epoch, including SMGs.   \citet{hua05}
  reported that 5\% of the LBGs at $z\sim 3$ in the sample of
  \citet{ste03} are detected at 24~\micron\, and
  they defined this sub-population as Infrared Luminous Lyman Break
  Galaxies (ILLBG).   ILLBGs have infrared properties very similar to
  sub--mm selected sources \citep{ega04,hua05,pope06}.   Huang et al.\
  used Spitzer [3.6]--[8.0] and [8.0]--[24] colors to argue that the
  24~\micron\ emission from the majority of ILLBGs is powered by
  intensive star formation.

  Observation of high redshift galaxies  with the Infrared
  Spectrograph (IRS, Houck et al. 2004) on board Spitzer permits the
  study of their 24~\micron\ emission mechanism.  Early IRS studies
  \citep{hou05, yan05} have focused on 24~\micron\ luminous
  (f$_{24}>$1 mJy) and optically faint sources. Most of the objects in
  both samples of \citet{hou05} and \citet{yan05}  are identified as
  dusty AGNs at $1.5<z<2.5$ whose spectra show power-law continua and
  strong silicate absorption but weak or absent emission features
  arising from the polycyclic aromatic hydrocarbon (PAH) molecules.
  \citet{lut05} observed two lensed SCUBA sources 
  at $z\sim2.8$ with the IRS. Both
  sources are so--called Hyper--luminous IR galaxies (HyLIRGs,
  L$_{8-1000\mu m} > 10^{13}$~L$_\odot$) with strong PAH features.

For galaxies at $z>2$, the Spitzer MIPS 24~\micron\ band detects
rest-frame mid--IR continuum emission.  At $z=2.1$ and $z=2.8$, the 6.2 and
7.7\micron\ PAH emission features respectively enter into the 24~\micron\ band.
Thus, a high redshift, 24~\micron\-selected-sample may
preferentially select objects with strong emission-line features in
the mid--IR bands.   We have been
studying just such a sample of LBGs, which have a
redshift distribution centered at $\langle$z$\rangle$=3.0
\citep{ste03}. 

  This letter presents the IRS spectrum and multiwavelength
  properties of one ILLBG at $z=3.01$ detected in the Extended Groth
  Strip (EGS) area.   The multi--wavelength data set
  from All-wavelength Extended Groth Strip International Survey
  (AEGIS, Davis et al. 2006)  including ACS images permits the study
  of Spectral Energy Distribution (SED) and morphology of this source.
  We adopt $\Omega_m=0.3$, $\Omega_{\Lambda}=0.7$, and $H_0=70$ km s$^{-1}$
  Mpc$^{-1}$.

\section{Multi-Wavelength photometry and Infrared Spectroscopy of an ILLBG}

  Most ILLBGs in the \citet{hua05} sample  have 24~\micron\
  flux densities of $\sim$0.1 mJy, too faint to permit efficient IRS
  observations.\footnote{see
  http://ssc.spitzer.caltech.edu/irs/documents/irs\_ultradeep\_memo.pdf}
  \citet{ash06} performed a wide-field deep u' and g' imaging survey
  with the MEGACAM on the MMT, covering the whole EGS and overlapping
  with deep Subaru R-band imaging and Spitzer/MIPS data.
  The goal of the MMT/Megacam survey is to obtain an
  ILLBG sample with higher 24~\micron\ flux density for IRS
  spectroscopic follow--up.   \citet{hua06} gave a detailed
  description of this sample.   Objects with strong AGN activity such
  as QSOs are excluded from the sample using combinations of
  [3.6]$-$[8.0] and [8.0]$-$[24] colors
  \citep{ivi04,hua05}. Spectroscopic redshifts for this sample are not
  yet available.   We selected the object that is the subject of this
  letter, EGS20 J1418+5236 (hereafter as E21, the 21st in the IRS target
  list) as a LBG with u'-g'$>$1.4, g'-R=0.39,
  and R=24.30, and with very strong 24~\micron\ emission, $f_{24}=0.62$~
  mJy.

  We obtained a spectrum for E21 with the IRS using the
  Low-resolution Long-wavelength (LL1) in staring mode, covering
  20--39~\micron.  We observed the source for 24 cycles of 120~s
  duration.  The total on-source integration time was 5851~s. We
  combined the reduced frames, subtracted residual background counts,
  and used the SMART package \citep{hig04} to extract calibrated
  one-dimensional spectra for both positive and negative beams.   The
  final IRS spectrum is an average of both beams.  
  The spectrum shows a significant
  detection of 6.2 and 7.7\micron\ PAH emission  features at $z=3.01\pm0.016$. 
This is consistent with a typical redshift for $u$--dropout--selected LBGs.
Given the presence of multiple emission features in the IRS spectrum
here, the redshift accuracy  is significantly better than that
reported for other $z>2$ sources in the literature. Spectra of other objects often have
low S/N, only a single emission feature, or featureless spectrum (e.g.,
Houck et al.\ 2005; Yan et al.\ 2005). 

\begin{figure}
\epsscale{1.0}
\plotone{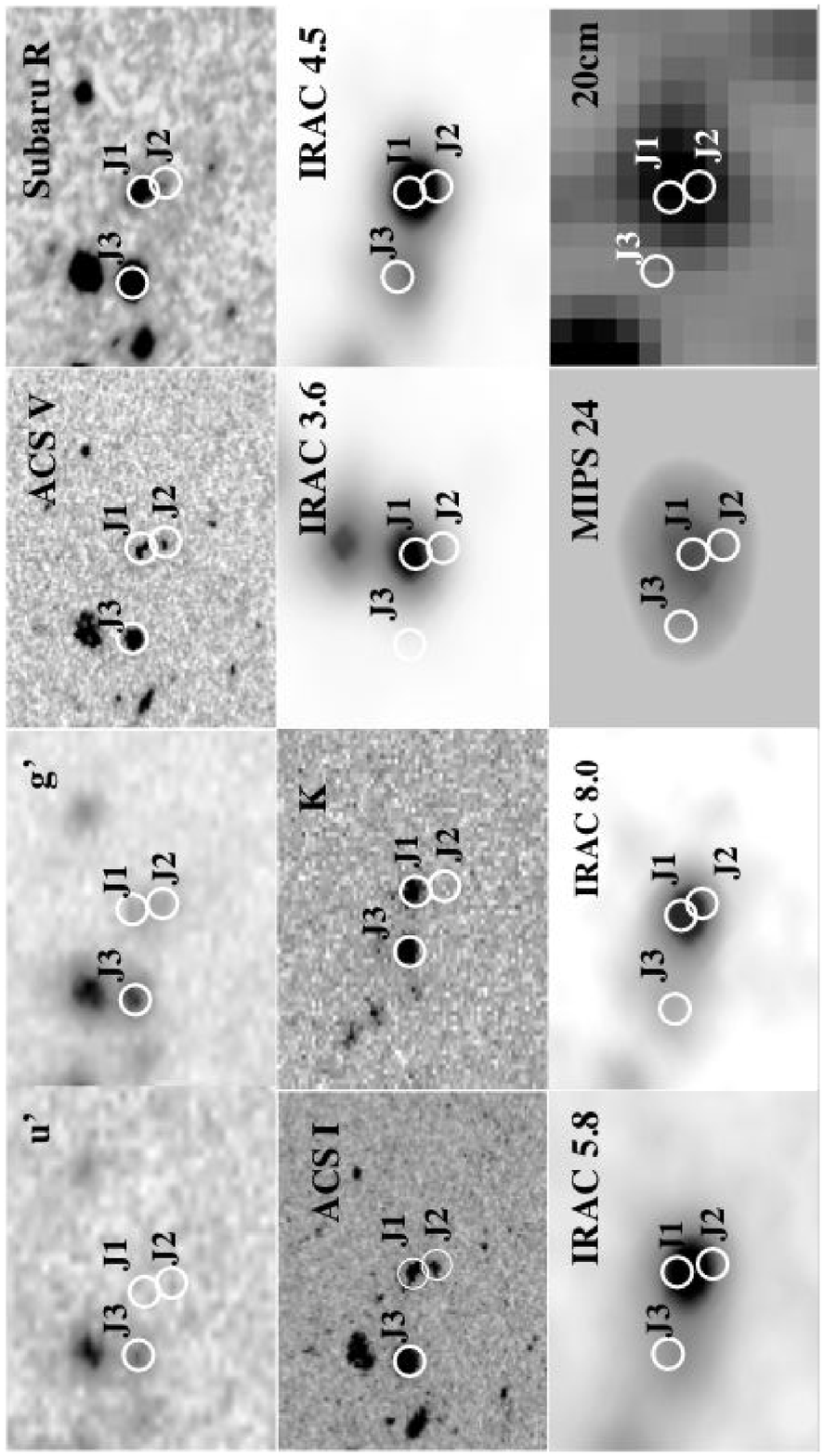}
\epsscale{1.0}
\caption{The multi-wavelength stamp images for EGS20 J1418+5236. The spatial resolution are
1\arcsec for the MMT/MEGACAM u' and g' images, 
0\arcsec6 for the Subaru R band image,
0\arcsec1 for the ACS V and I band images, 0\arcsec8 for the 
Palomar K-band image, 1\arcsec8-2\arcsec2
for the 4 IRAC bands, 5\arcsec for the MIPS 24 band, and 3\arcsec 
for the VLA 20cm image.
\label{f:stamp}}
\end{figure}

  The AEGIS multi-wavelength data set permits the study of E14
  from X-ray to 20 cm radio band. 
  While there is a single source in the MMT u', g', and
  Subaru R images, the HST/ACS images resolve this galaxy into two
  components, which we refer to as J1 and J2 (Figure~\ref{f:stamp}).  In
  addition,  another object, called J3, resides 3\arcsec\ to the
  east of J1 and J2, and could contaminate the Spitzer IR emission.
  Strong emission at u' from J3 suggests that it is at a much lower
  redshift, and using the full dataset we derive a photometric
  redshift of $z=0.79$.
J3  is not detected at 20~cm  (5$\sigma$ limit of
$f_{20cm}<$42~$\mu$Jy, compared to the 10$\sigma$ detection of the J1+J2
component with $f_{20cm}=$82~$\mu$Jy, Ivison et al. 2006). 
Therefore, we conclude that the J3
component does not contribute to the 24~\micron\ flux density or the
emission in the IRS spectrum. 

  Both J1 and J2 are so--called u' dropouts (e.g., Steidel et
  al.\ 2003) and show low surface--brightness features that
  connect to J1 and J2 in the ACS V and I band images. The
  two components are separated by 0\arcsec8 or 6.24 kpc at
  z=3.01. These facts suggest that  J1 and J2 belong to an
  interacting/merging system. The two components have different
  colors: J1 is red ($V_{AB}-I_{AB}$=0.97), and J2 is blue
  ($V_{AB}-I_{AB}$=0.65).  J1 is detected in both the J and K' bands, but
  J2 is not detected to $K_{limit,AB}=22.4$. J1 has
  $J_{AB}-K_{AB}$=0.88 (or 1.9 in the Vega system). 
  We argue that J1 is mainly responsible for the IRAC, MIPS 24~\micron,
  and radio flux densities and for the emission in the IRS spectrum.
  With $K_{AB}$=20.89 for J1 and $K_{limit,AB}=22.4$ for J2, J2 may
  contribute less than 11\% of the infrared
  emission of this system (see Figure~\ref{f:sed}) .

\section{The Infrared Properties of E21}

    Both the continuum and the 6.2~\micron\ PAH emission feature
    contribute to the 24~\micron\ flux density  ($f_{24}$=0.62 mJy) of
    E21.  We measure the continuum
    emission as 0.55 mJy in the observed wavelength range of
    20$<\lambda<$32.5~\micron, thus the contribution of the 6.2
    \micron\ PAH emission to the MIPS 24~\micron\ flux density is 0.07
    mJy.   Most ILLBGs have 24 \micron\ flux densities of
    $\sim$0.06--0.1 mJy \citep{hua05}.   If other ILLBGs have
    comparable emission line fluxes, then this line would contribute
    significantly to their 24~\micron\ flux densities.

  The power mechanism for  the mid-infrared emission of ILLBGs remains
  unclear.   PAH emission features are associated with intensive star
  formation \citep{rig99, lu03, lut05}. 
  \citet{rig99} suggest that  the 7.7\micron\
  feature-to-continuum ratio, L/C,  can be used to classify AGNs and
  starburst galaxies in ULIRG samples: L/C$>$1 for starburst galaxies
  and L/C$<$1 for AGNs. Though a few AGN dominated ULIRGs  and most
  type-II AGN could have L/C$>$1 \citep{lu03},  a large 7.7\micron\
  L/C value does at least suggest that intensive star formation is
  the dominant energy source.  \citet{pee04} argued that the
  6.2\micron\ PAH emission feature is a better indicator to separate
  starbursts and AGNs: QSO and type-I AGNs have typical
  feature-to-continuum ratio of 0.1, and starburst ULIRGs  and type-II
  AGNs  have a lower value in range of $0.5<L/C<2$.  The 6.2 and
  7.7\micron\ feature-to-continuum ratios for E21 are
  0.40 and 1.8 respectively, consistent with values of starburst
  ULIRGs.   Furthermore, the limit on X-ray emission from E21
  in the 200ks Chandra observation \citep{nan06}  is
  $L_{2-10Kev}<1.0\times10^{43} erg s^{-1}$, suggesting the absence of
  strong AGN activity.
  Two other
  high redshift SMGs with IRS spectroscopy are identified as AGN
  dominated ULIRGs with similar IR luminosities \textit{and} strong
  X-ray emission:  SMM J02399-0136 at z=2.8 \citep{lut05}, which much
  like this source has rather strong PAH emission; and CXO
  J141741.9+582823 at z=1.15 \citep{flo06}, which has no PAH emission
  features. Furthermore, even when AGN are detected in high--redshift
  SMGs based on their X--ray emission, Alexander et al.\ (2005) argue
  that substantial ongoing star formation dominates the IR emission in
  these objects.   Thus, we argue that intensive star formation
  is the dominant energy source  for the infrared emission of E21.
  The different types of IR--luminous, high--redshift
  sources may represent different stages of HyLIRGs: starburst
  dominated E21 with strong PAH and no X-ray emission;
  intermediate type SMM J02399-0136 and other SMGs with both PAH and
  detected X-ray emission; AGN dominated CXO J141741.9+582823 with no
  PAH but strong X-ray emission.

  Currently there is no sub-millimeter observation of this source. We
  use the 20cm radio flux and $f_{850\mu m}/f_{20cm}$-redshift
  relation \citep{car00,dun00,wan04a} to predict that E21
  has an 850\micron\ flux density of $10<f_{850\mu m}<17$ mJy.
  Similarity, we derive the rest-frame far-infrared luminosity using
  the FIR-radio luminosity relation \citep{con92,car00}:
\begin{equation}
L_{FIR}=4 \pi D_l^2 f_{20cm} 10^q (1+z)^{-(1+\alpha)}
\end{equation}
\begin{equation}
L_{8-1000\mu m}=1.92L_{FIR}
\end{equation}
where q=log($f_{FIR}/f_{20cm}$)=2.34,   and the radio spectral index
$\alpha=-0.8$.   For $f_{20cm}=0.082$ mJy at z=3.01, 
$L_{8-1000\mu m}$ is equal to
2$\times$10$^{13}$ L$_{\odot}$, qualifying  this source as a HyLIRG.

\begin{figure}
\plotone{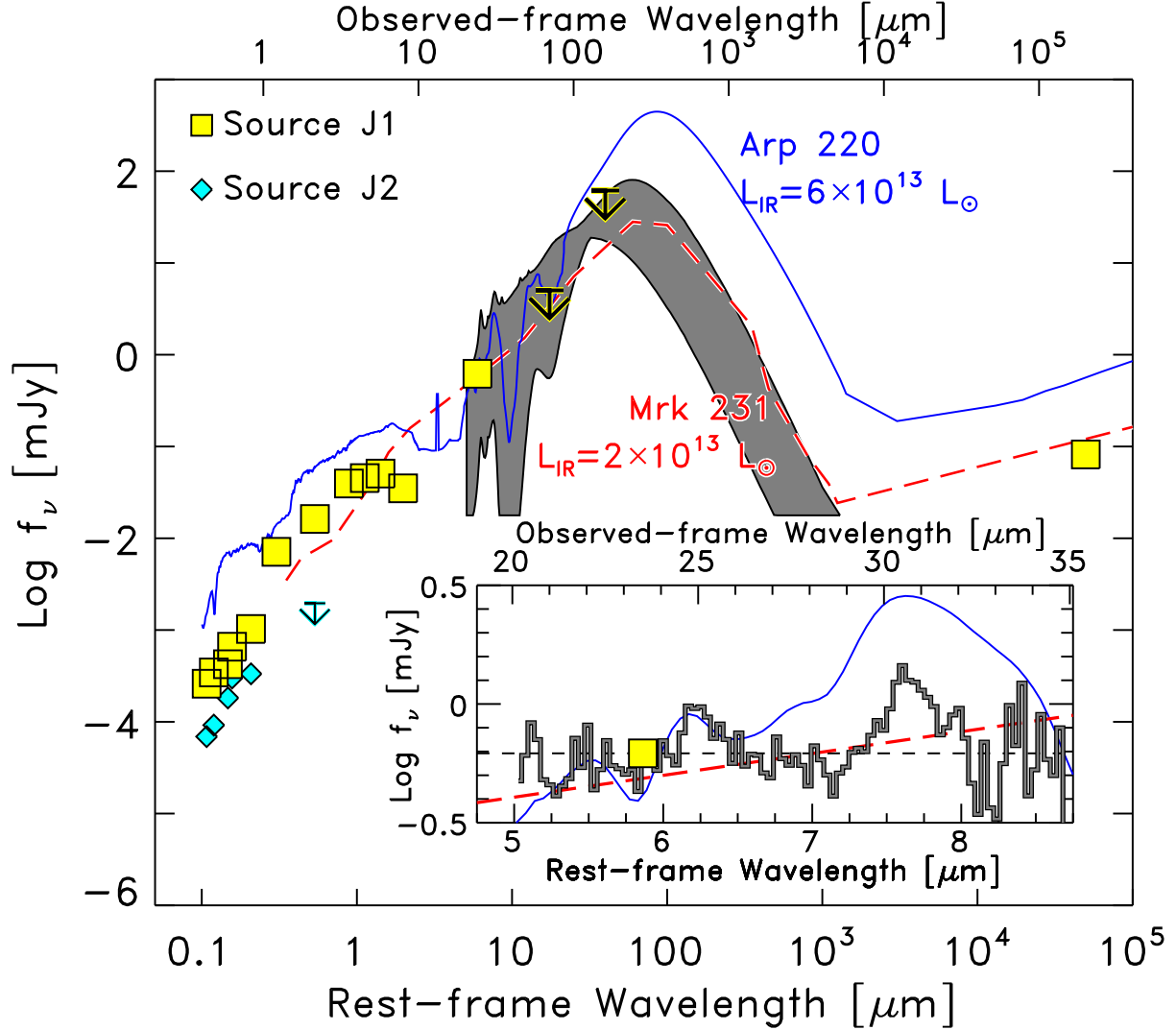}
\caption{The SEDs for both J1 and J2. The yellow boxes show the SED
  for J1 and cyan diamonds show the SED for J2.  Upper limits on the
  flux--density ($5\sigma$) are denoted by downward arrows.  The blue
  solid line shows the SED for Arp~220 (Chary \& Elbaz 2001) and the
  red dashed line shows the SED of Mrk~231, in both cases normalized
  to the flux density at 24~\micron.    J2 is not detected in either
  the J-- or   K--band images. Because J2 should not be redder than
  J1, we use the J1 and J2 K--band flux ratio (with the limiting flux
  for J2) to estimate that J2 should contribute less than 11\% to the
  IRAC  and MIPS 24~\micron\ photometry.   We use both the Arp220 and
  Mrk~231 SED models to estimate the total infrared luminosity of this
  source using the 24~\micron\ flux density. The Mrk~231 SED model
  with L$_{8-1000\mu m}$=2$\times$10$^{13}$ is in better agreement
  with the far--IR and Radio  flux  densities.   Similarity, the
  gray--shaded region shows the range of theoretical SEDs of
  Siebenmorgen \& Kruegel (2006) for starburst--powered IR--luminous
  galaxy models with L$_{8-1000\mu m}$ consistent with that estimated
  from the radio emission.   Note that these models have not been rescaled
  to fit the observed photometry, but instead we show the predicted flux
  densities for a galaxy at z=3.01. The inset plot shows the IRS spectrum.
  The horizontal dashed line shows the observed 24~\micron\ flux
  density (yellow box) and the other curves are the SEDs of Arp 220
  and Mrk 231 as for the main figure.
\label{f:sed}}
\end{figure}

Fitting the SED with various galaxy templates yields a similar IR
luminosity.   We normalize the SEDs of two local ULIRGs, Arp~220 and
Mrk~231,  to the 24~\micron\ flux density to predict the total infrared
luminosity for this source: 6$\times$10$^{13}$ L$_{\odot}$ and 
2$\times$10$^{13}$ L$_{\odot}$ respectively.
(Fig.~\ref{f:sed}). 
The Arp~220 SED model predicts a much higher far--IR 
and radio flux density than observed.
The Mrk231 SED provides a consistent interpretation to both the 24~\micron\ and
radio flux densities, as well as the 70 and 160~\micron\ flux limits.
Similarly, figure~\ref{f:sed} also compares the observed SED to
models with IR luminosity L$_{8-1000 \mu m}$=10$^{13.2-13.3}$~L$_\odot$
(i.e., consistent with that derived from the radio emission) from
Siebenmorgen \& Kruegel (2006).   
The models
broadly agree with the 24~\micron\ flux density (and the limits at 70
and 160~\micron) and provide further evidence that the SED is dominated by
emission from hot dust.    This is also consistent with strong
continuum emission in the IRS spectrum.   Therefore, given the
constraints from the empirical and theoretical models, we conclude
that hot dust dominates the far--IR emission in this object.

This source is very unlikely to be amplified due to gravitational
lensing.  Gravitational lensing is very inefficient for J3 being a
lensing galaxy at z=0.79. There are also no clusters or groups found
in the area.  Furthermore, the galaxy morphologies in the ACS image
and color difference of J1 and J2 do not support strong
lensing for this source.

The morphology of E21 is reminiscent of  those of local
ULIRGs.   Locally, about one third of ULIRGs have double nuclei
with a mean linear separation of 6.2~kpc
\citep[e.g.,][]{rig99,sur00}.   E21 appears to have
multiple components in the ACS images, see figure~\ref{f:stamp}.  The
projected separation of J1 and  J2 is 6.75~kpc.   
UV luminosity is a direct measurement for the relatively unobscured
massive star formation in ULIRGs.   Though ULIRGs are dusty, they
typically have strong UV emission \citep[e.g.,][]{sur00},
although this emission often stems from separate star--forming regions
rather than those that dominate the IR (and bolometric) emission
(e.g., Zhang et al.\ 2001; Wang et al.\ 2004b).   The U
($\lambda_{central}$=3414\AA) luminosities, $\nu L_{\nu}$,   for
ULIRGs \citep{sur00}  cover a  large range from 10$^9$ L$_{\odot}$ to
5$\times$10$^{11}$ L$_{\odot}$ with  an average of 2$\times$10$^{10}$
L$_{\odot}$.  At z$\sim$3, the observed J-band probes roughly the
rest-frame U band.   We use the J-band magnitude to estimate the
rest-frame U luminosity for EGS20 J1418+5236, $\nu
L_{\nu}(U)=3.5\times10^{11}$, roughly $\sim$1--2\% of the bolometric
emission.  This is consistent with its local IR--luminous galxay counterparts.

\section{Conclusions}
 
IRS spectra of EGS20 J1418+5236, a
$z=3.01$, u--dropout LBG in AEGIS show the 6.2 and 7.7\micron\ PAH emission features.
This is currently the
highest redshift galaxy where the PAH emission features have been
detected. This source has a total infrared luminosity of L$_{8-1000\mu
m}$=2$\times$10$^{13}$ L$_{\odot}$ inferred from both the 24~\micron\ and
radio flux densities.    Nearby galaxies with these IR luminosities that
are predominantly QSO/AGN, but given the presence of strong PAH
emission features and the lack of any evidence for an AGN, we conclude
that star--formation dominates the emission from this z=3 ILLBG.

The 24~\micron-to-radio flux ratio, strong IRS continuum emission, and
upper limits on the flux density at 70 and 160~\micron\ implies that
hot dust dominates the far--IR emission in this object, very similar
to Mrk 231 and model SEDs of HyLIRGs.
Interestingly, the theoretical models that best describe the shape and
relative strength of the mid--IR SED
require very high
extinction values ($A_V \gtrsim 70$~mag).  Because this object has
strong UV emission, we
suggest that this LBG contains a deeply embedded starburst in addition
to the relatively unobscured stars that dominate the rest--frame UV
optical emission.  This scenario may explain the general UV--IR
properties of the ILLBG population.   This example suggests that
HyLIRGs at high redshifts can appear as either ILLBGs, or SMGs, or
both. Larger samples of $z\sim 3$ galaxies with IRS spectroscopy and
measurements of their far-IR SED are clearly needed to confirm these
hypotheses.    

\acknowledgements

We wish to thank our collaborators on the AEGIS project for many
insightful conversations and much hard work.  
This work is based in part on observations made with the Spitzer Space
Telescope, which is operated by the Jet Propulsion Laboratory,
California Institute of Technology under a contract with NASA. Support
for this work was provided by NASA through an award issued by
JPL/Caltech. 



\clearpage

\clearpage
\begin{figure}
\plotone{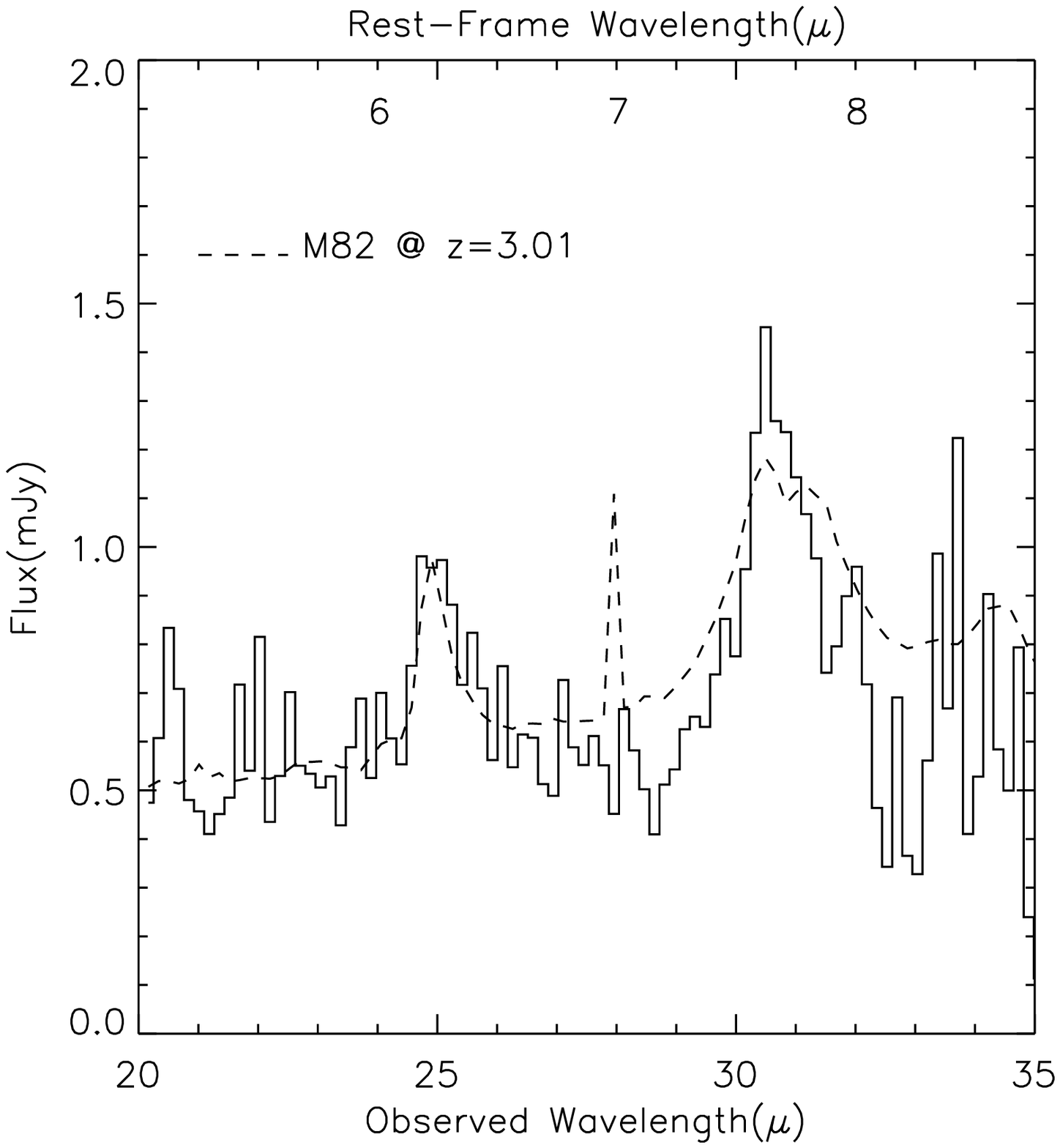}
\caption{The IRS spectrum of EGS20 J1418+5236. The dashed line is the
M82 SED shifted to z=3.01.The spectrum has
remarkably similar 6.2 and 7.7 \micron\ PAH
emission--feature strength and shape to those 
of M82.   We cross--correlated the IRS spectrum with that of M82 to derive a
redshift of $z=3.01\pm0.016$. The IRS spectrum at 
$\lambda>$32\micron\ becomes very
noisy, so the 8.2\micron\ PAH emission  feature  is not detected.
\label{f:irs}}
\end{figure}

\clearpage

\clearpage

\end{document}